\documentstyle[12pt,epsfig]{article}
\advance\hoffset by -7mm  
\makeatletter
\@addtoreset{equation}{section}
\makeatother

\renewcommand{\title}[1]{\null\vspace{25mm}

\noindent{\Large{\bf #1}}\vspace{10mm}

\noindent {\large By }}
\newcommand{\authors}[1]{\noindent{\large #1}\vspace{3mm}

}
\newcommand{\address}[1]{\noindent #1\vspace{5mm}

}
\renewcommand{\abstract}[1]{\vspace{19mm}

\noindent{\small{\em Abstract.} #1}\vspace{2mm}

} 

\begin{document}
\begin{flushright}
Z\"urich University Preprint\\
ZU-TH 4/96
\end{flushright}
\title{Asymptotic Behavior of the
Einstein-\kern -2.18ptYang-Mills-Dilaton\\[2mm] System
for a Closed Friedmann-Lemaitre Universe$\footnote{This work was
supported by the Swiss National Science Foundation.}$}
\authors{Reto Eggenschwiler, David Scialom and Norbert Straumann}
\address{Institute of Theoretical Physics, University of Z\"urich,
Winterthurerstrasse 190,\\CH-8057 Z\"urich, Switzerland}
\abstract{We study the coupled Einstein-Yang-Mills-Dilaton (EYMD) equations
for a Fried\-mann-Le\-mai\-tre universe with constant curvature $k=1$.
Our detailed analysis is restricted to the case where the dilaton
potential and the cosmological constant vanish. Also assuming a static gauge
field, we present analytical and numerical
results on the behavior of solutions of the EYMD equations.
For different values
of the dilaton coupling constant we analyze the phase portrait for
the time
evolution of the dilaton field and give the behavior of the scale
factor.
It turns out that there are no inflationary stages in this model.}
\section{Introduction}
For a long time the matter used in most cosmological
models was described in terms of ideal or viscous hydrodynamics. This is
certainly adequate at least back to the time of nucleosynthesis,
because---to a first approximation---we are then allowed to regard 
the matter as a mixture of pressure-free matter (baryons, WIMPS, etc...) and
radiation (photons, neutrinos). At very high energies the fluid
description has to be replaced by field theoretical models (and, when
approaching the Planck scale, by yet unknown theories of matter).
\hfil\vadjust{\vskip\parskip}\break\indent
With the success of gauge field theories, and especially in connection
with various proposed inflationary scenarios \cite{Linde},
field theoretical models became very popular in cosmology.
A description of topological defects, which could be
created in cosmological phase transition, requires also field
theoretical models which allow for spontaneous symmetry breaking.
\hfil\vadjust{\vskip\parskip}\break\indent
Inflationary stages appear more or less naturally in many models, but
most of those which lead to successful inflation are very
phenomenological in nature and are not based directly on
``fundamental'' field theories. Since inflation is a very attractive
general idea and for other reasons as well, it is important to study
the evolution of a large variety of cosmological models based on
interesting field theoretical matter descriptions. In addition to Yang-Mills
fields, the existence of a dilaton field is theoretically well
motivated on the basis of superstring theory \cite{Green}
and Kaluza-Klein theories \cite{Abbot}.
\hfil\vadjust{\vskip\parskip}\break\indent
In the present work we study the coupled Einstein-Yang-Mills-Dilaton
(EYMD) equations for a Friedmann-Lemaitre (Robertson-Walker)
space-time. The basic equations for this model lead to a dynamical
system of six first-order nonlinear ordinary differential equations,
which we discuss with the modern techniques of the theory of
dynamical systems. In view of the rather high dimension of phase space
this goal is not easy to achieve. In a first step we therefore analyze 
in detail only a restricted class of solutions. We do so by setting the
cosmological constant and the dilaton potential equal to zero, and
also require that the Yang-Mills field is static, which can be done 
consistently.
In this case the phase space becomes three-dimensional. We hope that
our analysis of this reduced system is at least methodically of some
interest. An extension of the discussion to the full system is planned.
\hfil\vadjust{\vskip\parskip}\break\indent
In a first step we compactify the phase space such that the
dynamical system can be extended to the enlarged space and determine
the critical points. With one exception, all critical points are
either located at
infinity of the original phase space or in the unphysical region.
We analyze the nature of these singular points and the
qualitative behavior of the flow in their vicinity. In general the
equilibrium points are hyperbolic. However, for special values of the
dilaton coupling constant there exists a critical point in the 
asymptotic physical region which
becomes non-hyperbolic. For this case we use the theory of normal
forms and apply the Poincar\'e-Dulac theorem to get rid of the
nonresonant nonlinear terms up to third order. Analytical formula
which are valid near the critical points are given in all cases.
\hfil\vadjust{\vskip\parskip}\break\indent
The asymptotic behavior provides one with the initial conditions for a
numerical study of the phase portrait. Special attention is given to
the value of the dilaton coupling constant which is obtained from
superstring theory. Unfortunately, and not unexpectedly, we do not
find any inflationary stages.  
\hfil\vadjust{\vskip\parskip}\break\indent
The
paper is organized as follows. In section~2 we derive the basic
equations.
Their analysis for the above-mentioned assumptions is performed in
section~3. We
analytically determine the asymptotic behavior of the solutions near
the singular points. Section~4
is devoted to numerical results. The behavior of the scale factor
as well as the phase portrait of the dilaton field are given. A short
summary concludes the paper.
\section{Basic Equations}
The metric of the Friedmann-Lemaitre universe with constant curvature
$k=1$ is given by
\begin{equation}
\label{metric}
  g = dt'^{2} - a^{2}(t')\,h = a^{2}(t)\,[dt^{2} - h],
\end{equation}
where $h$ is the standard metric on $S^{3}$ and $t$ is the conformal
time related to cosmic (proper) time by $dt'=a(t)\,dt$.

We work in the ``Einstein conformal frame'' in which the action for the
EYMD system is given by
\begin{equation}
\label{action}
  S=\frac{1}{4\pi}\int
  \left(-\frac{R}{4G}-\frac{\Lambda}{2G}+\frac{1}{2}(\partial\phi
  )^{2}-V(\phi)-\frac{e^{-2\kappa\phi}}{4}F^{2}\right)\,\eta\, ,
\end{equation}
where $F=dA + A \wedge A$, $\kappa$ denotes the dilatonic coupling
constant and $\eta$ is the volume form.
$A$ and $\phi=\phi (t)$ are the $SU(2)$ Yang-Mills potential and the
dilaton field, respectively.
Both are homogeneous and isotropic.

Since the group manifold of $SU(2)$ is $S^{3}$, there is a natural
ansatz for a
homogeneous isotropic $SU(2)$ gauge field involving a single function
of time~\cite{Gibbons}:
\begin{equation}
\label{ansatz}
  A = f(t)\,\Theta ,
\end{equation}
where $\Theta $ is the Maurer-Cartan form on $S^{3}$ regarded as
$SU(2)$.
Note that $t$
and $f$ must be considered dimensionless while the scale factor $a$
has the dimension of a length.
The canonical metric $h$ on $S^3$ can also be expressed in terms of
$\Theta$: If we decompose the $su(2)$-valued form according to
\begin{equation}
\Theta=\sum \theta^j \tau_j\, ,
\end{equation}
where $\tau_j=-i\sigma_j$ ($\sigma_j$ are the Pauli matrices), then the
$\theta^j$ are left invariant 1-forms and 
\begin{equation}
h=\delta_{ij}\, \theta^i\otimes \theta^j\, .
\end{equation}
A simple way to verify the normalization is to use the following
parameterization of $SU(2)$ in terms of $x=(x^1, x^2, x^3, x^4) \in
S^3$:
\begin{equation}
  U = x^{4} - x^{j}\tau _{j} =
  \left( \begin{array}{cc}
  x^{4} + ix^{3} & x^{2} + ix^{1}  \\
  -x^{2} + ix^{1} & x^{4} - ix^{3}
  \end{array} \right).
\end{equation}
The Maurer-Cartan form is then $U^{-1}dU$ and the $\theta^i$ are given
by
\begin{equation}
  \theta ^{i} = -\frac{1}{2}\mbox{tr}[\tau _{i}\,U^{-1}dU]\quad
  (i=1,2,3).
\end{equation}
We define an orthonormal basis of one-forms by
\begin{equation}
  \hat{\theta}^{0} := a\,dt,\; \hat{\theta}^{i} := a\,\theta ^{i}\quad
  (i=1,2,3).
  \label{basisforms}
\end{equation}
The metric~(\ref{metric}) then reads
\begin{equation}
   g = \hat{g}_{\mu\nu}\,\hat{\theta}^{\mu}\otimes
   \hat{\theta}^{\nu},\quad  \hat{g}_{\mu\nu}
   = \mbox{diag}(1,-1,-1,-1).
\end{equation}
Variation of the action with respect to the metric yields the
Einstein field
equations
\begin{equation}
  \label{fieldequations}
  \hat{G}_{\mu\nu} = 8\pi G\hat{T}_{\mu\nu} + \Lambda\hat{g}_{\mu\nu},
\end{equation}
with
\begin{eqnarray*}
  \hat{T}_{00} & = &
  \frac{1}{4\pi}\left\{\frac{\dot{\phi}^{2}}{2a^{2}}+V(\phi )+
\frac{3}{2a^{4}}[\dot{f}^{2}+4(f^{2}-f)^{2}]e^{-2\kappa\phi}\right\},\\
  \hat{T}_{ij} & = &
  \frac{1}{4\pi}\left\{\frac{\dot{\phi}^{2}}{2a^{2}}-V(\phi
  )+\frac{1}{2a^{4}}[\dot{f}^{2}+4(f^{2}-f)^{2}]e^{-2\kappa\phi}\right\}
  \delta_{ij},
\end{eqnarray*}
and $\hat{T}_{0i} = 0$. The (00) component of Einstein's
field equations is the constraint equation
\begin{equation}
  H^{2} + 1 = G\{\frac{1}{3}\dot{\phi}^{2}+\frac{2a^{2}}{3}V(\phi)
  +\frac{1}{a^{2}}[\dot{f}^{2} +4(f^{2} -
  f)^{2}]e^{-2\kappa\phi}\}+\frac{\Lambda}{3}a^{2},\label{eq1}
\end{equation}
with the Hubble parameter $H=\dot{a}/a$, where an overdot means the
derivative with respect to the conformal time $t$.
For the spatial components we have
\begin{equation}
   -[2\dot{H} + H^{2} + 1] = G\{\dot{\phi}^{2}-2a^{2}V(\phi)
   +\frac{1}{a^{2}}[\dot{f}^{2}
   +4(f^{2}-f)^{2}]e^{-2\kappa\phi}\}-\Lambda
   a^{2}.\label{eq2}
\end{equation}
The dilaton and the YM equation are obtained from variations of the
action with respect to $\phi$ and $f$, repectively,
\begin{eqnarray}
  \ddot{\phi}+2H\dot{\phi}+a^{2}\frac{\partial
  V}{\partial\phi}+\frac{3\kappa}{a^{2}}[\dot{f}^{2}-4(f^{2}-f)^{2}]e^{-2\kappa\phi}
  & = & 0, \label{eq3} \\
  \ddot{f}-2\kappa\dot{\phi}\dot{f}+4f(2f-1)(f-1) & = & 0. \label{eq4}
\end{eqnarray}
The system is fully determined by the independent
equations~(\ref{eq1}), (\ref{eq3})
and (\ref{eq4}). In fact, one can (easily) show that Eq.~(\ref{eq2}) is
a consequence of the others. This is a reflection of the Bianchi
identities.

From Eq.~(\ref{eq4}) we see that the static gauge field solutions are
given \mbox{by $f=0,\,1/2,\,1$.}
Only the case $f=1/2$ is of interest, because $f=0$
corresponds to a vanishing gauge field ($A=0$) and $f=1$ is a pure
gauge ($A=\Theta $).

A systematic study of the complete system is rather involved.
We confine ourselves to the special case
of a static gauge field $f=1/2$ and choose in addition $V=0$ and
$\Lambda =0$. The set of differential equations (\ref{eq1})-(\ref{eq4})
then reduces
to a three-dimensional dynamical system. In order to analyze this
reduced system,
we use the same procedure as in~\cite{David}. We determine the singular
points~\cite{Arrow} (see page 15), including those lying at infinity of the 
phase space,
and then find analytically the asymptotic behavior for the solutions
near these
points.
\section{Asymptotic Behavior}
Setting $f=1/2$, $V=0$, $\Lambda = 0$
and introducing a ``proper'' time coordinate $\tilde t$ according to
\begin{equation}
\label{newtime}
  d\tilde{t}=\frac{dt}{a(t)},
\end{equation}
we obtain the transformed system of basic equations
\begin{eqnarray}
  \phi _{\tilde{t}\tilde{t}} & = & -\tilde{H}\phi
  _{\tilde{t}}+\frac{3\kappa}{4}
  e^{-2\kappa\phi}, \label{flow1} \\
  \tilde{H}_{\tilde{t}} & = & \tilde{H}^{2}-G[\frac{2}{3}\phi
  ^{2}_{\tilde{t}}+\frac{1}{4}e^{-2\kappa\phi}],
  \label{flow2}
\end{eqnarray}
\begin{equation}
\label{flow3}
  \tilde{H}^{2}+a^{2} = G[\frac{1}{3}{\phi}^{2}_{\tilde
  {t}}+\frac{1}{4}e^{-2\kappa\phi}],
\end{equation}
where $\tilde{H}=a_{\tilde{t}}/a$ and the subscript $\tilde{t}$
denotes the
derivative with respect to this variable.

Equations~(\ref{flow1}) and (\ref{flow2}) form a three-dimensional
dynamical system
in the phase space $\phi ,\;\phi _{\tilde{t}},\;\tilde{H}$. We now
introduce
the following dimensionless variables:
\begin{eqnarray*}
  \tilde{t} & \rightarrow & \eta = \sqrt{G}\,\tilde{t}, \\
  \kappa & \rightarrow & \tilde{\kappa}=\mbox{\small
  $\sqrt{\frac{3}{G}}$}\,\kappa ,\\
  \phi & \rightarrow & x=\mbox{\small $\sqrt{\frac{G}{3}}$}\,\phi ,\\
  \phi _{\tilde{t}} & \rightarrow & y= \mbox{\small
  $\frac{\;1}{\!\!\!\sqrt{3}}$}\,\phi_{\tilde{t}}
  , \\
  \tilde{H} & \rightarrow & z=\frac{\tilde{H}}{\sqrt{G}}, \\
  a & \rightarrow & \tilde{a}=\frac{a}{\sqrt{G}}.
\end{eqnarray*}
Then Eqs.~(\ref{flow1}) and (\ref{flow2}) become
\begin{eqnarray}
  x_{\eta} & = & y, \\
  y_{\eta} & = &
  -yz+\frac{\tilde{\kappa}}{4}e^{-2\tilde{\kappa}x},\label{mech} \\
  z_{\eta} & = & z^{2}-2y^{2}-\frac{1}{4}e^{-2\tilde{\kappa}x},
\end{eqnarray}
and the constraint~(\ref{flow3}) takes the form
$y^{2}+\frac{1}{4}e^{-2\tilde{\kappa}x}-z^{2}=\tilde{a}^{2}$ with
$z=\tilde{a}_{\eta}/\tilde{a}$.

For the special case $\tilde{\kappa}=0$, these
equations are equivalent to a one-dimensional
mechanical system for $a(\eta)$ with the Lagrange function
$L=T-V=\tilde{a}^{2}_{\eta}-(\tilde{a}^{4}-\tilde{a}^{2}/4)$.
Indeed, Eq.~(\ref{mech}) immediately implies the relation
$y=k/\tilde{a}$,
where $k$ is an arbitrary constant. From the constraint equation we
obtain the corresponding
conserved ``energy''
$E=T+V=\tilde{a}^{2}_{\eta}+(\tilde{a}^{4}-\tilde{a}^{2}/4)=k^{2}\geq
0$.
The solution with respect to the conformal time is
\begin{eqnarray*}
  \tilde{a}(t)& = &
  \frac{\sqrt{2}}{4}\,\left[1+\sqrt{1+64k^{2}}\,\sin{(2t-\lambda
  )}\right]^{\frac{1}{2}}, \\
z(t) & = & \frac{\sqrt{2}}{4}\,\frac{\sqrt{1+64k^{2}}\,\cos{(2t-\lambda
)}}{(1+\sqrt{1+64k^{2}}\,\sin{(2t-\lambda
)})^{\frac{1}{2}}}, \\
x(t) & = & \frac{1}{2}\,\ln{\left[\frac{\tan{(t-\lambda
/2)}+\sqrt{1+64k^{2}}-8k}{\tan{(t-\lambda/2)}+\sqrt{1+64k^{2}}+8k}\right]}+\mbox{const.}\qquad
t\in (0,
\pi/2+\lambda),
\end{eqnarray*}
where $\lambda $ is defined by
\begin{displaymath}
 \lambda :=\arcsin{\left(\frac{1}{\sqrt{1+64k^{2}}}\right)}.
\end{displaymath}

If $\tilde{\kappa}\not= 0$, we use instead of $x$ the function
\begin{equation}
  u=\frac{e^{-\tilde{\kappa}x}}{2}>0. \label{u}
\end{equation}
Our system then takes the final form
\begin{eqnarray}
  u_{\eta} & = & -\tilde{\kappa}uy, \nonumber \\
  y_{\eta} & = & -yz+\tilde{\kappa}u^{2}, \label{fsystem} \\
  z_{\eta} & = & z^{2}-u^{2}-2y^{2}, \nonumber
\end{eqnarray}
with the constraint
\begin{equation}
\label{fconstraint}
  u^{2}+y^{2}-z^{2}=\tilde{a}^{2}.
\end{equation}
where we recall that $z$ is the Hubble parameter with respect to $\eta$,
\begin{equation}
\label{fHubble}
  z=\frac{\tilde{a}_{\eta}}{\tilde{a}}.
\end{equation}

The system of equations~(\ref{fsystem})-(\ref{fHubble}) possesses the
following
symmetries:
\begin{tabbing}
\hskip 3cm \= a) $u\rightarrow -u$\quad\quad \= b) \= $\eta\rightarrow -\eta
$\quad\quad \= c) \= $\tilde{a}\rightarrow -\tilde{a} $\quad\quad \= d)
\= $y\rightarrow -y$ \\
\> \> \> $y\rightarrow -y$ \> \> \> \> $\tilde{\kappa}\rightarrow
-\tilde{\kappa}$ \\
\> \> \> $z\rightarrow -z$
\end{tabbing}
According to a) the plane $u=0$ is a two-dimensional invariant
subspace of the phase space; it separates
solutions with $u>0$ from those with $u<0$. Trajectories can reach this
plane asymptotically, but they never penetrate it. We are not
interested in solutions
with $u\equiv 0$; for these the change of variables given by Eq.~(\ref{u})
would
no longer be valid.

\begin{figure}[!ht]
\begin{center}
\epsfig{file=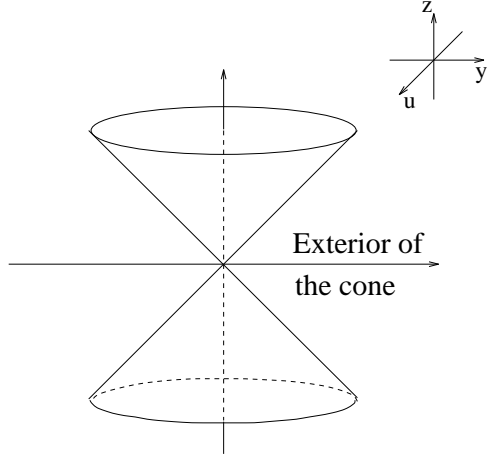}
\end{center}
\caption{Physical domain.}\label{fig:cone}
\end{figure}
Equations~(\ref{u}) and (\ref{fconstraint}) imply that the
trajectories of our model lie in the region of phase space exterior
to the cone
$u^{2}+y^{2}-z^{2}=0$ where $u>0$ (see Fig.~1).

We
also note that the sections of the trajectories lying in the upper half
of the phase
space ($z>0$) correspond to expansion ($H>0$) and
those in
the lower ($z<0$) to contraction ($H<0$). When a 
trajectory intersects
the plane $z=0$ ($H=0$), the scale factor
$\tilde{a}$ attains
its maximum, because $z_{\eta}=-(u^{2}+2y^{2})$ is strictly
negative.
\subsection{The Singular Point at the Origin}
Looking for equilibrium states of the dynamical system~(\ref{fsystem}) we
immediately see
that the origin $(u,y,z)=(0,0,0)$ is the only singular point not lying
at infinity.
However, this critical point is not of much interest, because there
are no trajectories originating from it which extend to the exterior of 
the cone nor does any trajectory (coming
from the exterior of the cone) ever reach the origin: Suppose there
would be a trajectory
emerging from the origin $(u,y,z)=(0,0,0)$. Along this,
$z=\tilde{a}_{\eta}/\tilde{a}$
has to be negative, since $z_{\eta}=-(y^{2}+\tilde{a}^{2})$ is strictly
negative.
It follows that if $\tilde{a}_{\eta}>0\; (\tilde{a}_{\eta}<0)$ along
the trajectory,
then $\tilde{a}$ must be negative (positive). This is a contradiction
because $\tilde{a}$ vanishes at the origin. With similar arguments one
can prove
that no  trajectory can reach the origin from the exterior of the
cone.

The repelling behavior of the singular point $(u,y,z)=(0,0,0)$ is
nicely illustrated
by considering the case where $\tilde{\kappa}=1$. Then the plane $z=-y$
is
an invariant subspace and the phase portrait of the trajectories lying
on this plane
is given in Fig.~2.
Furthermore, we can give the explicit solution. On the invariant plane
$z=-y$ our system
of Eqs.~(\ref{fsystem}) reduces to
\begin{eqnarray}
  u_{\eta } & = & uz, \\
  z_{\eta } & = & -z^{2}-u^{2}, \label{explicit}
\end{eqnarray}
and the constraint equation becomes $u^{2}=\tilde{a}^{2}$.
Eq.~(\ref{explicit}) is equivalent to
the equation of motion of a one-dimensional mechanical
system with the Lagrange function $L=T-V=\tilde{a}_{\eta
}^{2}/2-\tilde{a}^{4}/4$.
The
corresponding conserved ``energy'' is $E=T+V=\tilde{a}_{\eta
}^{2}/2+\tilde{a}^{4}/4\geq
0$.
The solution with respect to the conformal time is
\begin{eqnarray*}
  \tilde{a}(t)& = & u(t)
	       = (2\sqrt{E}\,\sin{(\sqrt{2}\,t)})^{\frac{1}{2}}, \\
  z(t) & = & -y(t) 
	       = 
	      \frac{\sqrt{E}\,\cos{\sqrt{2}\,t}}{(\sqrt{E}\,\sin{(\sqrt{2}
	      \,t)})^{\frac{1}{2}}},\\
  x(t) & = &
  -\frac{1}{2}\ln{\left[8\sqrt{E}\,\sin{(\sqrt{2}\,t)}\right]}\quad\qquad
  t\in (0,\pi/\mbox{\scriptsize $\sqrt{2}$}).
\end{eqnarray*}
In the limit $t\rightarrow 0$, the scale factor $\tilde{a}$ vanishes
and the equation of state tends to $\epsilon=p$ (stiff matter), where
$\epsilon =
\hat{T}_{00}$
and $p=\frac{1}{3}(\hat{T}_{11}+\hat{T}_{22}+\hat{T}_{33})$. At
$t=\sqrt 2\frac{\pi}{2}$, $\tilde{a}$ attains its maximum and 
the equation of
state becomes $\epsilon=3p$ which corresponds, for ordinary matter, to a
extremely relativistic gas.
\begin{figure}[!h]
\begin{center}
\epsfig{file=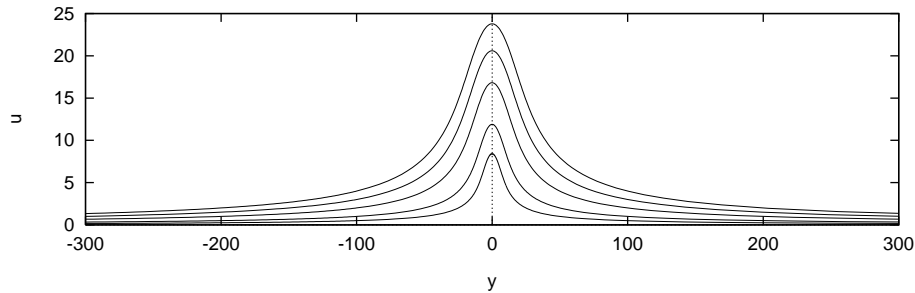}
\end{center}
\caption{Phase portrait of the invariant plane $z=-y$.}\label{nfig1}
\end{figure}
\subsection{The Singular Points at Infinity}
In order to find and investigate the singular points at infinity
it is convenient
to compactify the phase space and complete it by the infinitely distant
boundary \mbox{$u^{2}+y^{2}+z^{2}=\infty $.} We do so by going over from the
Cartesian coordinates
$u,y,z$ to the spherical coordinates
\begin{equation}
\label{spherical}
  u=r\cos\varphi\sin\vartheta ,\;\; y=r\sin\varphi\sin\vartheta ,\;\;
  z=r\cos\vartheta ,
\end{equation}
and by performing a subsequent transformation of the radius according to
\begin{equation}
\label{rho}
  r=\frac{\rho}{1-\rho}\quad (0\leq\rho <1).
\end{equation}
Also introducing a new time $\tau $ defined by
\begin{equation}
\label{tau}
  d\eta = (1-\rho )\,d\tau ,
\end{equation}
and using the variables $\rho ,\vartheta ,\varphi ,\tau $, our system
(\ref{fsystem}) assumes the form
\begin{eqnarray}
  \rho _{\tau} & = & \rho ^{2}(1-\rho )\cos\vartheta [\cos (2\vartheta
  )-2\sin ^{2}\vartheta\sin^{2}\varphi],
  \nonumber \\
  \vartheta _{\tau} & = & -\rho\sin\vartheta\cos (2\vartheta )
  [1+\sin^{2}\varphi],
  \label{csystem} \\
  \varphi _{\tau} & = & \rho\cos\varphi [\tilde{\kappa}\sin\vartheta -
  \sin\varphi\cos\vartheta ].\nonumber
\end{eqnarray}
For $0\leq\rho <1$, Eqs.~(\ref{fsystem}) are equivalent to the
system~(\ref{csystem}),
but the latter admits a smooth continuation to the boundary $\rho =1$.
Thus, in the variables $0\leq\rho\leq 1$, $0\leq\vartheta\leq\pi $,
$0\leq\varphi\leq
2\pi $, the phase space of the system~(\ref{csystem}) is compact. Each
point of the
original phase space is mapped along its radius vector to a certain
point
inside the unit ball, while the points $u^{2}+y^{2}+z^{2}=\infty
$ at infinity
are mapped to its surface $\rho =1$. From
Eqs.~(\ref{csystem})
it is readily seen that on this surface the system has several singular
points,
their number and nature depending on the value of $\tilde{\kappa}$. Due
to the \mbox{symmetry
transformation d)} it is sufficient to consider only positive values
of $\tilde{\kappa}$ (i.e. $\tilde{\kappa}>0$). There are at least six
singular points
which will be denoted by (see Fig.~3)
\begin{displaymath}
\begin{array}{lll}
  P(\vartheta =0), & K_{1}((\vartheta _{1},\varphi _{1})=(\pi
  \!/4,3\pi\!/2)), &
  K_{2}((\vartheta _{2},\varphi _{2})=(\pi\!/4,\pi\!/2)),\\
  P'(\vartheta '=\pi), & K'_{1}((\vartheta _{1}',\varphi
  _{1}')=(3\pi\!/4,\pi\!/2
  )), & K'_{2}((\vartheta _{2}',\varphi _{2}')=(3\pi \!/4,3\pi\!/2)).
\end{array}
\end{displaymath}
For $\tilde{\kappa}<1$ we have the four additional points
\begin{displaymath}
\begin{array}{lll}
  K_{3}((\vartheta _{3},\varphi _{3})=(\pi\!/4,\arcsin
  \tilde{\kappa})), & K_{4}(
  (\vartheta _{4},\varphi _{4})=(\pi\!/4,\pi -\arcsin
  \tilde{\kappa})),\\
  K'_{3}((\vartheta _{3}',\varphi _{3}')=(3\pi\!/4,-\arcsin
  \tilde{\kappa})), &
  K'_{4}((\vartheta_{4}',\varphi_{4}')=(3\pi\!/4,\pi
  +\arcsin \tilde{\kappa})).
\end{array}
\end{displaymath}
In the limit $\tilde{\kappa}\rightarrow 1$, these coincide 
with $K_{2}$ and $K'_{2}$
respectively.
The symmetries a) and b) imply that there are only four
essentially different points: $P,\;K_{1},\;K_{2}$ and $K_{3}$. Since
the singular point $P$
at the north pole lies in the {\it interior} part of the cone, it is not of
physical interest.

Before discussing the remaining singular points we would like to make
the following
remark: For the three points $K_{i}\hskip0.2cm (i=1,2,3)$ there exists
a separatrix which goes from $K_{i}$ straight down to the origin
$(0,0,0)$, moving
on the surface of the cone. In fact, Eqs.~(\ref{csystem}) have admit
solutions of
the form $(\vartheta (\tau ),\varphi (\tau))\equiv (\vartheta
_{i},\varphi _{i})
$, where $(\vartheta _{i},\varphi _{i})$ denotes the angular
coordinates of $K_{i}$,
along which we have $\rho _{\tau}\leq 0$. This is not
contradictory to
the statements made in section~3.1, because the scale factor
$\tilde{a}$ vanishes along
these separatrices.
\begin{figure}
\begin{center}
\epsfig{file=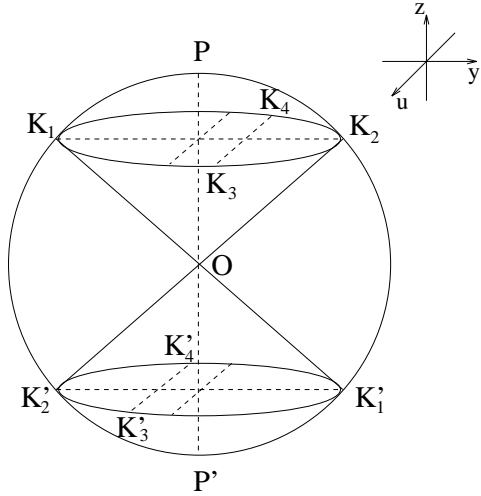}
\end{center}
\caption{Singular points at infinity.}\label{fig:ball}
\end{figure}
\subsubsection{The Singular Point K$_{1}$}
In order to study the asymptotic behavior of the solutions near $K_{1}$
we first
perform a shift, of the variable, defined by $\delta\rho = \rho
-1,\;\delta\vartheta = \vartheta
- \vartheta _{1}$ and \mbox{$\delta\varphi = \varphi -\varphi _{1}$}.
In
these new coordinates, the linearization of Eqs.~(\ref{csystem}) at
$K_{1}$ gives
\begin{equation}
\label{linearization}
\left( \begin{array}{c} \delta\rho \\ \delta\vartheta
\\ \delta\varphi        \end{array}
\right)_{\tau}=
\left( \begin{array}{ccc} \lambda _{\rho} & 0 & 0 \\
			  0 & \lambda _{\vartheta} & 0 \\
			  0 & 0 & \lambda _{\varphi} \end{array}
			  \right)
\left( \begin{array}{c} \delta\rho \\ \delta\vartheta
\\ \delta\varphi        \end{array}
\right),
\end{equation}
with the strictly positive eigenvalues $\lambda
_{\rho}=\sqrt{2}/2,\;\lambda _{\vartheta}=2\sqrt{2}$
and \mbox{$\lambda _{\varphi}=\sqrt{2}/2\, (1+\tilde{\kappa})$}.
Hence, the singular point $K_{1}$ is hyperbolic~\cite[p. 69]{Arrow} and
the leading
terms near $K_1$ are given by
\begin{eqnarray}
(\delta\rho )(\tau ) & = & -(\delta\rho )_{o}\, e^{\lambda
_{\rho}\tau},\quad  (
\delta\vartheta )(\tau )=(\delta\vartheta )_{o}\, e^{\lambda
_{\vartheta}\tau},\nonumber\\
(\delta\varphi )(\tau ) & = & (\delta\varphi )_{o}\, e^{\lambda
_{\varphi}\tau}\quad
(\tau\rightarrow -\infty ),\label{abK1}
\end{eqnarray}
with arbitrary integration constants $(\delta\rho
)_{o}>0,\;(\delta\vartheta )_{o}\geq
0$ and $(\delta\varphi )_{o}>0$. Integration of Eq.~(\ref{tau}) leads
to
the relation $e^{\lambda _{\rho}\tau}=\frac{\lambda
_{\rho}}{(\delta\rho )_{o}}\,\eta$,
i.e., emergence from $K_{1}$ corresponds to increasing $\eta $,
starting from
$\eta =0$.

For the Cartesian coordinates $(u,y,z)$ we obtain (indicating only the
leading terms as
$\eta\rightarrow 0^{+}$)
\begin{eqnarray}
u(\eta ) & = & (\delta\varphi )_{o}\left[\frac{\lambda
_{\rho}}{(\delta\rho
)_{o}}\right]^{1+\tilde{\kappa}}\,\eta^{\tilde{\kappa}},\nonumber
\\
y(\eta ) & = & -\frac{1}{\eta},\label{asym1} \\
z(\eta ) & = & \frac{1}{\eta}.\nonumber
\end{eqnarray}
From Eq.~(\ref{fHubble}) it is readily seen that near $K_{1}$ the scale
factor follows
the law $\tilde{a}(\eta )=C\,\eta $, where $C>0$ is a constant, fixed by
the constraint~(\ref{fconstraint}).

Hence, to the point $K_{1}$ there corresponds an initial cosmological
singularity at
a certain finite time which can always be chosen at $t=0$. Near $K_{1}$
we have
\begin{equation}
H=\frac{1}{2t},\quad\dot{\phi}=-\sqrt{\frac{3}{4G}}\,\frac{1}{t},\quad\phi
=-\sqrt{\frac{3}{4G}}\,\ln
(\frac{t}{t_{o}}),
\end{equation}
where $t_{o}>0$ is an arbitrary constant. Emergence from $K_{1}$
corresponds to
increasing $t$, beginning from $t=0$. The effective equation of state
near $K_{1}$
tends to $\epsilon =p$ (i.e., stiff matter).
\subsubsection{The Singular Point K$_{2}$}
In order to study the asymptotic behavior of the solutions near $K_{2}$
we proceed
in a similar manner. Linearization at $K_{2}$ again leads to
Eq.~(\ref{linearization})
with the same eigenvalues $\lambda _{\rho}$ and $\lambda _{\vartheta}$,
but
with the changed eigenvalue \mbox{$\lambda
_{\varphi}=\sqrt{2}/2\,(1-\tilde{\kappa})$}.
We suspect that in this case the behavior may depend on the value
of $\tilde{\kappa}$. For $\tilde{\kappa}< 1$ all eigenvalues are again
strictly
positive. Hence, the singular point $K_{2}$ is hyperbolic and the
leading terms
in the expansion around $K_2$ are given by Eq.~(\ref{abK1}) with 
arbitrary constants
$(\delta\rho )_{o}>0,\;(\delta\vartheta )_{o}\geq
0$ and $(\delta\varphi )_{o}<0$. Integration of
Eq.~(\ref{tau}) leads to the relation $e^{\lambda
_{\rho}\tau}=\frac{\lambda _{\rho}}{(\delta\rho )_{o}}\,\eta$,
i.e. emergence from $K_{2}$ corresponds to \mbox{increasing $\eta$,}
beginning from $\eta =0$.

For the Cartesian coordinates $(u,y,z)$ we obtain in
leading order as
$\eta\rightarrow 0^{+}$
\begin{eqnarray}
u(\eta ) & = & -(\delta\varphi )_{o}\left[\frac{\lambda
_{\rho}}{(\delta\rho
)_{o}}\right]^{1-\tilde{\kappa}}\,\eta^{-\tilde{\kappa}},\nonumber
\\
y(\eta ) & = & \frac{1}{\eta},\label{asym2} \\
z(\eta ) & = & \frac{1}{\eta}.\nonumber
\end{eqnarray}

Near $K_{2}$ the scale factor follows the law $\tilde{a}(\eta )=C\,\eta
$, where
$C>0$ is a constant fixed by the constraint~(\ref{fconstraint}).
Hence, if $\tilde{\kappa}<1$, $K_2$ corresponds to an initial
cosmological singularity at a finite time
$t=0$, say. Near $K_{2}$ we have
\begin{equation}
H=\frac{1}{2t},\quad\dot{\phi}=\sqrt{\frac{3}{4G}}\,\frac{1}{t},\quad\phi
=\sqrt{\frac{3}{4G}}\,\ln
(\frac{t}{t_{o}}),
\end{equation}
where $t_{o}>0$ is an arbitrary constant. Emergence from $K_{2}$
corresponds to
increasing $t$, starting from $t=0$. The equation of state tends to
$\epsilon =
p$ (i.e., stiff matter).

If $\tilde{\kappa}>1$, the singular point $K_{2}$ is also hyperbolic,
however, $\lambda_{\varphi}$
is strictly negative. The outgoing trajectories necessarily
lie on the invariant plane $u=0$ (i.e: $(\delta\varphi )_{o}=0$) and
are thus not of
interest.

The case $\tilde{\kappa}=1$ is more involved because the singular point
$K_{2}$
is non-hyperbolic. Terms of higher order become important. Up to third
order the
expansion in a sufficiently small neighborhood around $K_{2}$ is
\begin{eqnarray}
(\delta\rho )_{\tau} & = & \frac{\sqrt{2}}{2}\,\delta\rho +
\sqrt{2}\,\delta\rho^{2}
+ \frac{3\sqrt{2}}{2}\,\delta\rho\delta\vartheta +
\frac{\sqrt{2}}{2}\,\delta\rho^{3}
\nonumber\\ & & \hspace{70pt}
-\frac{9\sqrt{2}}{4}\,\delta\rho\delta\vartheta^{2}
+ 2\sqrt{2}\,\delta\rho ^{2}\delta\vartheta - \frac{\sqrt{2}}{2}
\,\delta\rho\delta\varphi ^{2},\nonumber\\
(\delta\vartheta )_{\tau} & = & 2\sqrt{2}\,\delta\vartheta +
2\sqrt{2}\,\delta\vartheta^{2}+2\sqrt{2}\,\delta\rho\delta\vartheta
\label{thirdorder}\\ & &  \hspace{70pt}
-\frac{7\sqrt{2}}{3}\,\delta\vartheta
^{3}+2\sqrt{2}\,\delta\rho\delta\vartheta^{2}-\sqrt{2}\,\delta\vartheta\delta\varphi^{2},\nonumber
\\ & & \nonumber\\
(\delta\varphi )_{\tau} & = & -\sqrt{2}\,\delta\vartheta\delta\varphi -
\frac{\sqrt{2}}{4}\,\delta\varphi^{3}
- \sqrt{2}\,\delta\rho\delta\vartheta\delta\varphi.\nonumber
\end{eqnarray}
We can now use the Poincar\'{e}-Dulac theorem~\cite[p. 191]{Arnold} to
classify
the nonlinear terms in Eq.~(\ref{thirdorder}) into resonant and
non-resonant ones. For
the non-resonant terms of order $n$ there is a polynomial change of
coordinates
of degree $n$ so that they are transformed into terms of at least
order $n+1$.
This is not the case for the resonant terms. The polynomial change of
coordinates
is found by solving the so-called homological equation. Performing the
polynomial
change of coordinates on Eq.~(\ref{thirdorder}), the non-resonant terms
of degree $2$ and $3$ become of higher order and are thus neglected.
Retaining only
terms up to third order, we obtain the set of equations
\begin{eqnarray}
\frac{dv_{\rho}}{d\tau} & = &
\frac{\sqrt{2}}{2}\,v_{\rho}-\frac{\sqrt{2}}{2}\,v_{\rho}v_{\varphi}^{2},\nonumber\\
 & & \nonumber\\
\frac{dv_{\vartheta}}{d\tau} & = &
2\sqrt{2}\,v_{\vartheta}-\sqrt{2}\,v_{\vartheta}v_{\varphi}^{2},\label{Dulac}\\
 & & \nonumber\\
\frac{dv_{\varphi}}{d\tau} & = &
-\frac{\sqrt{2}}{4}\,v_{\varphi}^{3},\nonumber
\end{eqnarray}
where $(\delta\rho ,\delta\vartheta ,\delta\varphi
)=(v_{\rho},v_{\vartheta},v_{\varphi})$
+ polynomes of degree $2$ or higher in
$(v_{\rho},v_{\vartheta},v_{\varphi})$.
The minus sign in the last equation implies that for
$\tilde{\kappa}=1$
there are no trajectories which emanate from $K_{2}$ into the exterior
of the
cone. Indeed, Eqs.~(\ref{Dulac}) can be solved analytically. We obtain
\begin{eqnarray*}
  v_{\varphi }(\tau) & = & \pm\, \frac{1}{\sqrt{(v_{\varphi
  o})^{-2}+\frac{\sqrt{2}\,\tau}{2}}}\qquad\quad\tau\in
  (-\sqrt{2}\,(v_{\varphi o})^{-2},\infty ), \\ & & \\
  v_{\rho}(\tau ) & = & \frac{v_{\rho o}}{(v_{\varphi
  o})^{2}}\,v_{\varphi}^{2}(\tau )\,e^{\frac{\sqrt{2}}{2}\,\tau},
  \\ & & \\
  v_{\vartheta}(\tau) & = & \frac{v_{\vartheta o}}{(v_{\varphi
  o})^{4}}\,v_{\varphi}^{4}(\tau )\,e^{2\sqrt{2}\,\tau},
\end{eqnarray*}
where $v_{\rho o}, v_{\vartheta o}$ and $v_{\varphi o}\neq 0$ are
arbitrary constants.
The $+\, (-)$ sign corresponds to the case $v_{\varphi o}>0$
($v_{\varphi o}<0)$.
Hence, for non-vanishing constants $v_{\rho o}$ and $ v_{\vartheta o}$,
we have
\begin{displaymath}
  v_{\varphi}(\tau )\rightarrow 0,\quad |v_{\rho}(\tau )|\rightarrow
  \infty,\quad
  |v_{\vartheta}(\tau )|\rightarrow \infty\qquad
  (\tau\rightarrow\infty ).
\end{displaymath}
\subsubsection{The Singular Point K$_{3}$}
The singular point $K_{3}$ only appears for $\tilde{\kappa}<1$.
Linearization at
$K_{3}$ leads to
\begin{equation}
\label{linearization3}
\left( \begin{array}{c} \delta\rho \\ \delta\vartheta
\\ \delta\varphi        \end{array}
\right)_{\tau}=
\left( \begin{array}{ccc} \mu _{\rho} & 0 & 0 \\
			  0 & \mu _{\vartheta} & 0 \\0 &
			  \mbox{\small
			  $\sqrt{2}\,\tilde{\kappa}\,\sqrt{1-\tilde{\kappa}^{2}}$}&
			  \mu_{\varphi}\end{array}\right)
\left( \begin{array}{c} \delta\rho \\ \delta\vartheta
\\ \delta\varphi        \end{array}
\right),
\end{equation}
with the eigenvalues $\mu _{\rho}=\sqrt{2}/2\,\tilde{\kappa}^{2},\;\mu
_{\vartheta}=\sqrt{2}\,(1+\tilde{\kappa}^{2})$ and
\mbox{$\mu_{\varphi}=-\sqrt{2}/2\,(1-\tilde{\kappa}^{2})$.}
The point $K_{3}$ is thus hyperbolic and the leading terms near
it are given by
\begin{eqnarray}
(\delta\rho )(\tau ) & = & -(\delta\rho )_{o}\, e^{\mu
_{\rho}\tau},\quad
(\delta\vartheta )(\tau
)=(\delta\vartheta )_{o}\, e^{\mu _{\vartheta}\tau},\nonumber
\\ (\delta\varphi )(\tau ) & = & \mbox{\small
$\frac{\sqrt{2}\,\tilde{\kappa}\,\sqrt{1-\tilde{\kappa}^{2}}}{\mu
_{\vartheta}-\mu_{\varphi}}$}\,(\delta\vartheta )_{o}\,
e^{\mu_{\vartheta}\tau}\quad  (\tau\rightarrow -\infty ),\label{abK3}
\end{eqnarray}
with arbitrary constants $(\delta\rho )_{o}>0$ and $(\delta\vartheta
)_{o}>0$.
Note that the general solution of~(\ref{linearization3}) would be given
by \[(\delta\varphi
)(\tau )=\mbox{\small
$\frac{\sqrt{2}\,\tilde{\kappa}\,\sqrt{1-\tilde{\kappa}^{2}}}{\mu_{\vartheta}-\mu_{\varphi}}$}\,(\delta\vartheta
)_{o}\,
e^{\mu_{\vartheta}\tau}+(\delta\varphi
)_{o}\,e^{\mu_{\varphi}\tau}.\]
However, since
$\mu _{\varphi}$ is negative the constant $(\delta\varphi )_{o}$ must
vanish.

Integration of Eq.~(\ref{tau}) leads to the relation $e^{\mu
_{\rho}\tau}=\frac{\mu_{\rho}}{(\delta\rho )_{o}}\,\eta$,
i.e., emergence from $K_{3}$ corresponds
to increasing $\eta $, starting from $\eta =0$.
In terms of the Cartesian coordinates $(u,y,z)$ we obtain
\begin{eqnarray}
u(\eta ) & = &
\tilde{\kappa}^{-2}\,\sqrt{1-\tilde{\kappa}^{2}}\,\frac{1}{\eta}-
\sqrt{(1-\tilde{\kappa}^{2})/2}\nonumber\\ & & \hspace{30
pt}+\,\tilde{\kappa}^{-2}\,\sqrt{1-\tilde{\kappa}^{2}}\;\frac{3-\tilde{\kappa}^{2}}{3+\tilde{\kappa}^{2}}\,(\delta\vartheta
)_{o}\left[\frac{\mu _{\rho}}{(\delta\rho
)_{o}}\right]^{\frac{\mu_{\vartheta}}{\mu_{\rho}}}\,\eta^{\frac{\mu_{\vartheta}}{\mu_{\rho}
}-1},\nonumber \\
y(\eta ) & = &
\tilde{\kappa}^{-1}\,\frac{1}{\eta}-\frac{\sqrt{2}}{2}\,\tilde{\kappa}+\tilde{\kappa}^{-1}\,\frac{5-\tilde{\kappa}^{2}}{3+\tilde{\kappa}^{2}}\,(\delta\vartheta
)_{o}\left[\frac{\mu _{\rho}}{(\delta\rho )_{o}}\right]^{\frac{\mu
_{\vartheta}}{\mu_{\rho}}}\,\eta^{\frac{\mu_{\vartheta}}{\mu_{\rho}}-1},\label{asym3}
\\
z(\eta ) & = &
\tilde{\kappa}^{-2}\,\frac{1}{\eta}-\frac{\sqrt{2}}{2}-\tilde{\kappa}^{-2}\,(\delta\vartheta
)_{o}\left[\frac{\mu _{\rho}}{(\delta\rho
)_{o}}\right]^{\frac{\mu_{\vartheta}}{\mu_{\rho}}}\,\eta^{\frac{\mu_{\vartheta}}{\mu_{\rho}}-1},\nonumber
\end{eqnarray}
with $\frac{\mu _{\vartheta}}{\mu
_{\rho}}=2(1+\tilde{\kappa}^{-2})>4$.
From Eq.~(\ref{fHubble}) it is readily seen that near $K_{3}$ the scale
factor
behaves like $\tilde{a}(\eta )=C\,\eta ^{1/\tilde{\kappa}^{2}}$, where $C>0$
is a constant, fixed by the constraint~(\ref{fconstraint}).

The point $K_{3}$ therefore corresponds to an initial cosmological
singularity at
a finite time $t=0$, say. Near $K_{3}$
we have
\begin{equation}
H=(1+\tilde{\kappa}^{2})^{-1}\,\frac{1}{t},\quad\dot{\phi}=\frac{\tilde{\kappa}}
{1+\tilde{\kappa}^{2}}\,\sqrt{\frac{3}{G}}\,\frac{1}{t},\quad\phi
=\frac{\tilde{\kappa}}{1+\tilde{\kappa}^{2}}\,\sqrt{\frac{3}{G}}\,\ln
(\frac{t}{t_{o}}),
\end{equation}
where $t_{o}>0$ is an arbitrary constant. Emergence from $K_{3}$
corresponds to
increasing $t$, starting from $t=0$.
\section{Numerical Results}
The asymptotic behavior of the solutions near the singular points
provides one with the initial
conditions for solving the differential equations numerically. We will
recover the repelling ($K_{1},K_{2},K_{3}$) and attracting 
($K'_{1},K'_{2},K'_{3}$)
characteristics of the singular points, as well as their dependence on
the
value of $\tilde{\kappa}$. Special emphasis is given to the value
$\tilde{\kappa}=\sqrt{3}$ which is predicted by a model obtained from
superstring theory~\cite{Green,George}.
\subsection{Emergence from K$_{1}$}
The asymptotic behavior~(\ref{asym1}) of the solutions near $K_{1}$
provides proper
initial conditions at a given time $\eta=\eta _{o}$. The constraint is
taken
into account by fixing the constant $C$ at the time $\eta _{o}$. We can
now produce
plots for different values of $\tilde{\kappa}$ and $k:=(\delta\varphi
)_{o}\left[\frac{\lambda_{\rho}}{(\delta\rho
)_{o}}\right]^{1+\tilde{\kappa}}$.

\noindent{\bf $\tilde{\kappa}=1/2:$}
Fig.~4 shows the behavior of the scale factor $\tilde{a}$ for different
values of the
constant $k$. The phase portrait of the dilaton field is given in
Fig.~5.
Since $\tilde{\kappa}<1$, each singular point 
\begin{figure}[!h]
\begin{center}
\epsfig{file=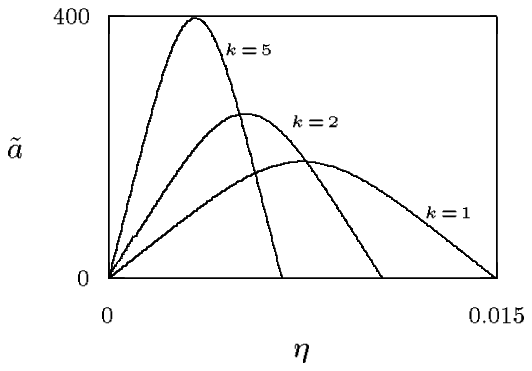}
\epsfig{file=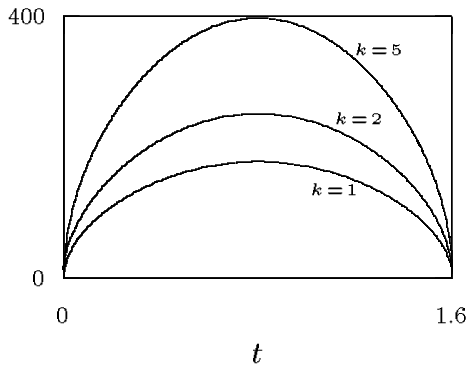}
\end{center}
\caption{The scale factor with respect to the time $\eta $ and the
conformal \mbox{time $t$} respectively.}\label{fig:1}
\begin{center}
\epsfig{file=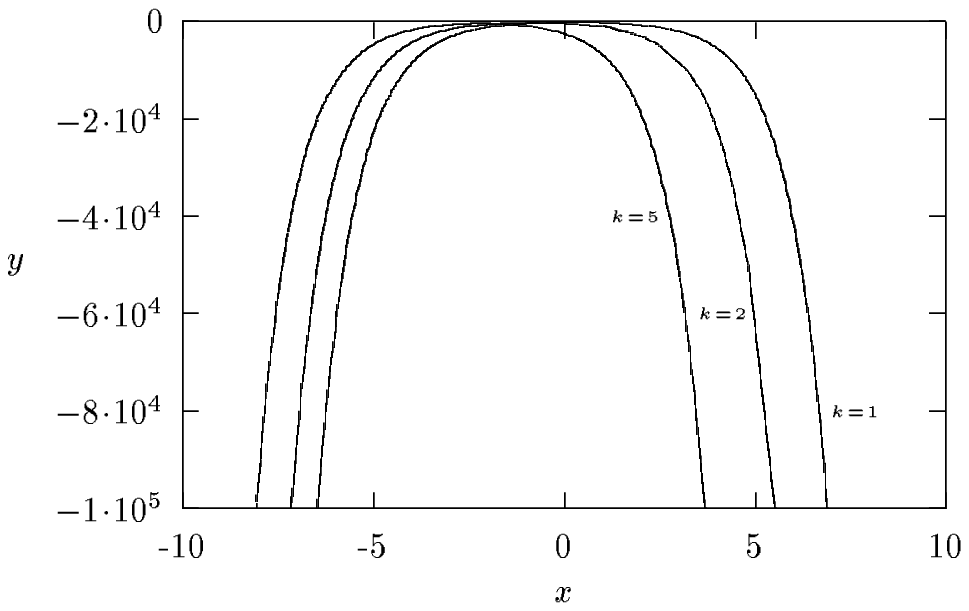}
\end{center}
\caption{Phase portrait of the dilaton field.}\label{fig:5}
\begin{center}
\epsfig{file=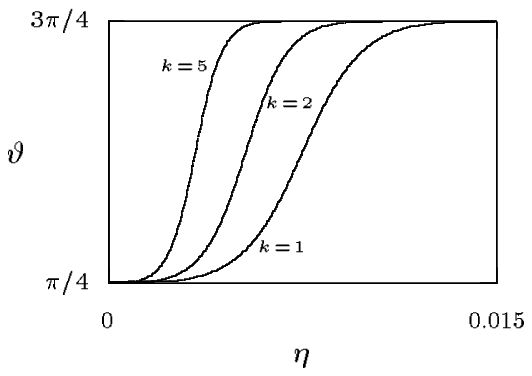}
\epsfig{file=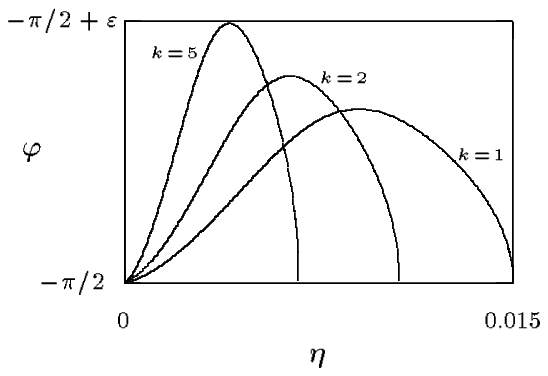}
\end{center}
\caption{The behavior of $\vartheta $ and $\varphi $ with respect to
$\eta$.
The value
of $\varepsilon$ is $\varepsilon = 8\cdot 10^{-4}$. The
trajectories are therefore close to the plane $u=0$.}\label{fig:2}
\end{figure}
$K'_{1},K'_{2}$ or
$K'_{3}$ is attractive.
From Fig.~6 it can be seen that the plots actually correspond to
trajectories
which end at $K'_{2}$. We have also computed numerically 
$\vartheta$ and $\varphi$ as a function of time for all subsequent
cases. This enable us to identify the points at which the trajectories
end. These plots can be found in the Diploma thesis of one of
us~\cite{regg}.

\noindent{\bf $\tilde{\kappa}=\sqrt{3}:$}
Fig.~7 shows the behavior of the scale factor $\tilde{a}$ for different
values of the
constant $k$. The phase portrait of the dilaton field is given in
Fig.~8.
Since for $\tilde{\kappa}\geq 1$ only $K'_{1}$ is an attractive point,
we expect
that the plots correspond to trajectories which end at $K'_{1}$. This
is confirmed by our numerical computation.
\begin{figure}[!h]
\begin{center}
\epsfig{file=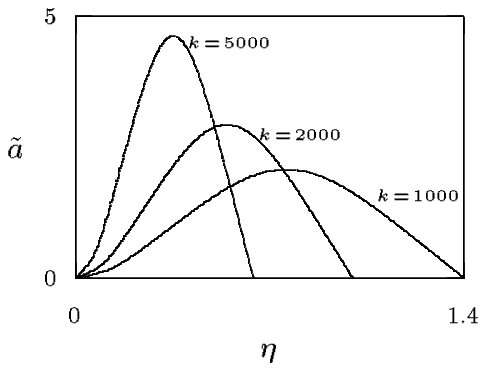}
\epsfig{file=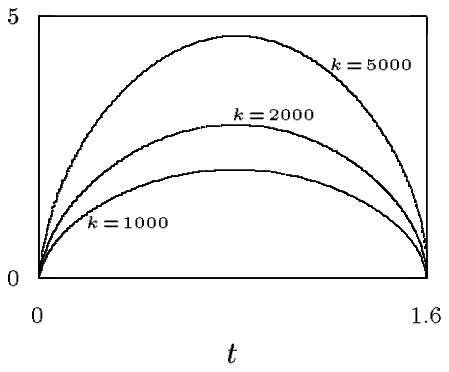}
\end{center}
\caption{The scale factor with respect to the time $\eta $ and the
conformal \mbox{time $t$} respectively.}\label{fig:6}
\begin{center}
\epsfig{file=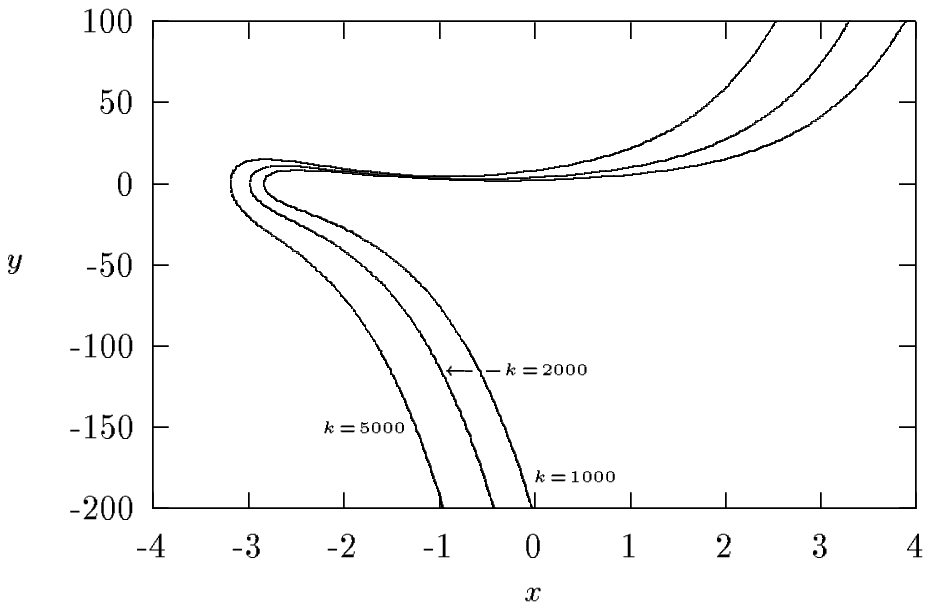}
\end{center}
\caption{Phase portrait of the dilaton field.}\label{fig:10}
\end{figure}
\subsection{Emergence from K$_{2}$}
We proceed in a similar manner. Emergence from $K_{2}$ only appears if
$\tilde{\kappa}<1$.
The constant $k$ is defined by $k:=-(\delta\varphi
)_{o}\left[\frac{\lambda_{\rho}}{(\delta\rho
)_{o}}\right]^{1-\tilde{\kappa}}$ (see Eq.~(\ref{asym2})).

\noindent{\bf $\tilde{\kappa}=1/2:$}
Fig.~9 shows the behavior of the scale factor $\tilde{a}$ for
different values
of the constant $k$. The phase portrait of the dilaton field is given
in Fig.~10. The
plots correspond to trajectories which end at the singular point
$K'_{1}$.
\begin{figure}[!t]
\begin{center}
\epsfig{file=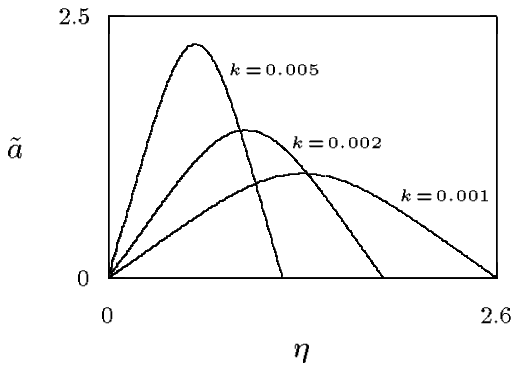}
\epsfig{file=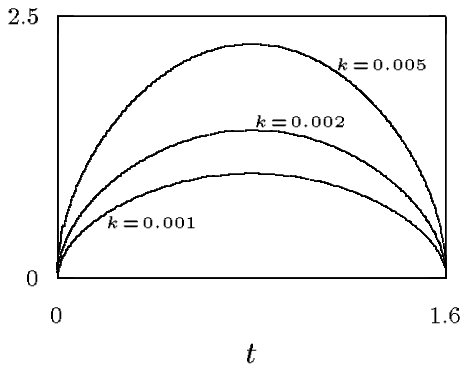}
\end{center}
\caption{The scale factor with respect to the time $\eta $ and the 
conformal \mbox{time $t$} respectively.}\label{fig:11}
\begin{center}
\epsfig{file=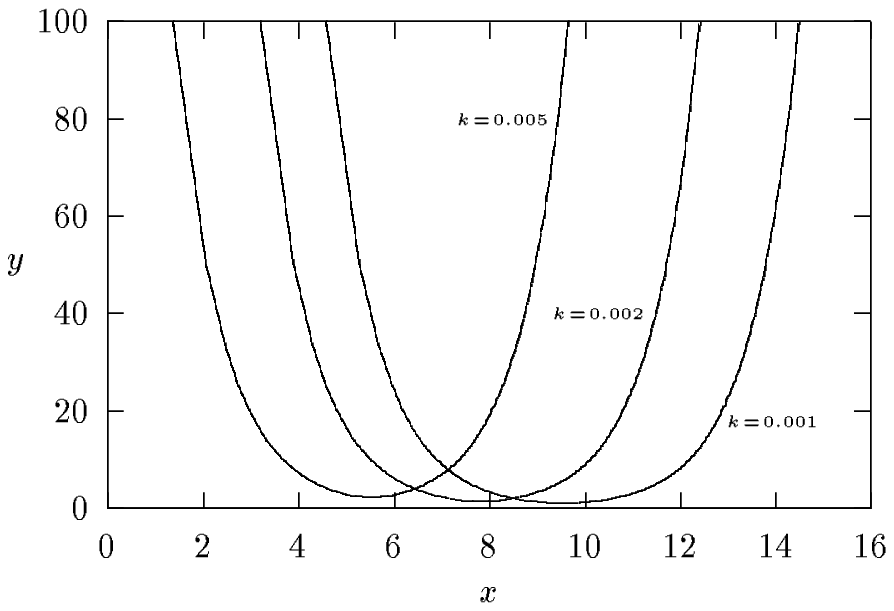}
\end{center}
\caption{Phase portrait of the dilaton field.}\label{fig:15}
\end{figure}
\subsection{Emergence from K$_{3}$}
The singular point $K_{3}$ only exists if $\tilde{\kappa}<1$. We define
the constant
$k$ by $k:=(\delta\vartheta )_{o}\left[\frac{\mu _{\rho}}{(\delta\rho
)_{o}}\right]^{\frac{\mu_{\vartheta}}{\mu_{\rho}}}$
(seeEq.~(\ref{asym3})).
\begin{figure}[!h]
\begin{center}
\epsfig{file=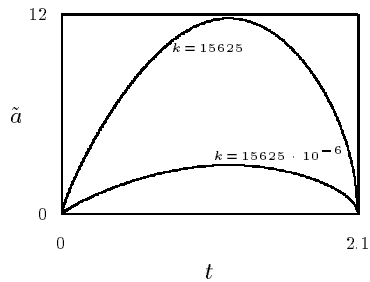}
\end{center}
\caption{The scale factor with respect to the conformal time
$t$.}\label{fig:16}
\end{figure}

\noindent{\bf $\tilde{\kappa}=1/2:$}
Fig.~11 shows the behavior of the scale factor $\tilde{a}$ for
different values
of the constant $k$. The phase portrait of the dilaton field is given
in Fig.~12. The plots correspond to trajectories
which end at
$K'_{1}$.
\begin{figure}[!h]
\begin{center}
\epsfig{file=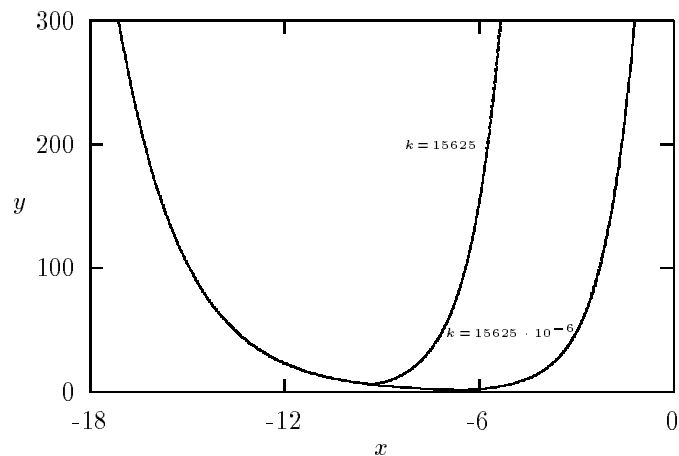}
\end{center}
\caption{Phase portrait of the dilaton field.}\label{fig:20}
\end{figure}
\section{Concluding Remarks}
For the special case of a static gauge field and
$V=\Lambda =0$ we have performed a complete analysis of the coupled 
EYMD equations for a
Friedmann-Lemaitre universe with constant curvature $k=1$ (closed
model). For this situation,
we did {\it not} recover any inflationary stages. It might therefore be
interesting to extend
our discussion to the more general case of a dynamical gauge field,
perhaps with
a non-vanishing cosmological constant $\Lambda $ and/or a dilaton
potential
$V$. This is, however, a considerably more involved task, because 
we encounter then
six coupled first-order nonlinear differential equations.

The analysis of the coupled EYMD equations for a Friedmann-Lemaitre
universe with
constant curvature $k=0$ (flat model) and $k=-1$ (open model) are
also of interest,
since related considerations~\cite{David,Bel} show that
inflationary
stages occur for trajectories which come close enough to one of the
separatrices
found for $k=0$.
\vskip 2\baselineskip
\noindent{\bf Acknowledgments}
\vskip \baselineskip
\noindent We are grateful to Othmar Brodbeck, George Lavrelashvili and Mikhail
Volkov for
very useful and clarifying discussions, and to Marcus Heusler for
careful reading of the manuscript.

\end{document}